\documentstyle[aps,twocolumn]{revtex}
\begin{document}
\draft
\author{Bao-Sen Shi and Akihisa Tomita}
\address{Imai Quantum Computation and Information Project, NEC Tsukuba Laboratories,\\
ERATO, Japan Science and Technology Corportation (JST)\\
Fundamental Research Laboratories, NEC, 34 Miyukigaoka, Tsukuba, Ibaraki,\\
305-8501, Japan}
\title{Schemes for generating W state of paths and W state of polarization photons}
\maketitle

\begin{abstract}
In this paper, we give two very simple schemes to produce two kinds of W
states, one kind is path W state with one photon and the other is
multiphoton photon polarization W state. These schemes just need a common
commercial multiport fiber coupler and single photon sources, they are
feasible by current technologies.
\end{abstract}


Entanglement is the key physical resource in most quantum information
process, e. g., quantum teleportation [1], quantum key distribution [2],
quantum computation[3] and so on. In multiparticle case, it was shown that
there exists two inequivalent classes of entangled states, namely
Greenberger-Horne-Zeilinger (GHZ) [4]\ state and W state [5], where, for
example in three-particle case, $\left| GHZ\right\rangle =\frac 1{\sqrt{2}}%
[\left| 000\right\rangle +\left| 111\right\rangle ]$ and $\left|
W\right\rangle =\frac 1{\sqrt{3}}[\left| 001\right\rangle +\left|
010\right\rangle +\left| 100\right\rangle ]$ , they can not been converted
to each other even under stochastic local operations and classical
communication. Recently, Cabello considered a set of Bell inequality to show
some differences between the violation of local realism exhibited by GHZ
state and W state.[6] One obvious application of general W state is
telecloning [7, 8]. If we choose the particular W state $\left| \Psi
_{clone}\right\rangle =\sqrt{\frac 23}\left| 100\right\rangle -\sqrt{\frac 16%
}\left| 010\right\rangle -\sqrt{\frac 16}\left| 001\right\rangle $, then we
can get precisely a Bu\v {z}ek-Hillery cloning [9] from the sender to two
receivers, provided that the results are averaged over the four possible
measurement outcomes of Bell state by sender. In Ref. [10], we give a scheme
by which W state can be used to realize the teleportation of an unknown
state probabilisticly. The teleportation of the entangled state and dense
coding by W\ state are discussed in Ref. [11]. In this paper, we consider
how to produce a W state. The GHZ state has been produced in laboratory
recently using Spontaneous-Parametric-Downcoversion. [12] How about W state?
In Ref. [13], authors present a quantum electrodynamics scheme, by which, a
multiatom W state can be produced. The Heisenberg model was used to produce
three-atom or four-atom W state in Ref. [14]. Very recently, Zou et. al [15]
present a scheme by which a four-photon or a three-photon W state can be
generated by linear optical elements. In their scheme, besides many linear
optical elements, a maximal entangled state source and single photon source
are needed. The setup seems complicated. In this paper, we give two very
simple schemes, by which, two kinds of W state can be produced very easily.
The first kind is the path W\ state, the character of the first kind W state
is that the total number of particle is just one. The successful probability
is 100\%. The second is the multiphoton polarization W state, we can get it
probabilisticly with postselection. In these two schemes, just a common
commercial multiport fiber coupler and single photon source are needed.
Contrast to the scheme [15], no maximal entangled source and other linear
optic elements are needed. So, they are very simple and easy to do in
practice. Furthermore, by these schemes, not only a three-mode or a four
mode W state (first kind) and a three-photon or a four-photon W state
(second kind), but also an arbitrary multimode (first kind) and multiphoton
(second kind) W state can be produced in principle, so our schemes are more
general than the scheme in Ref. [15]. In the follow, we give these schemes
to produce two kinds of different W states respectively. Firstly, we discuss
how to generate the first kind of W state.

We take how to produce the state $\left| W\right\rangle =\sqrt{\frac 13}%
[\left| 001\right\rangle +\left| 010\right\rangle +\left| 100\right\rangle
]_{123}$ as an example, where, subscripts 1, 2, 3 refer to the different
three space modes, and $\left| 0\right\rangle $ means vacuum state, $\left|
1\right\rangle $ means one photon state. In order to do it, what we need is
just a 3$\times $3 symmetric fiber coupler (tritter) if we do it in fiber
system. A symmetric 3$\times $3 fiber can be described by a unimodulat
matrix. A standard form of the tritter matrix $T$, in which the first column
and the first row are real, is given by [16]

\begin{equation}
T=\sqrt{\frac 13}\left[ 
\begin{array}{ccc}
1 & 1 & 1 \\ 
1 & \exp (i\frac{2\pi }3) & \exp (i\frac{4\pi }3) \\ 
1 & \exp (i\frac{4\pi }3) & \exp (i\frac{2\pi }3)
\end{array}
\right] .
\end{equation}
If the input state to the tritter is $\Psi _{in}=(1,0,0)_{123}$, where, 1,
2, 3 refer to three input ports of tritter, one photon is in input 1, the
inputs to 2 and 3 are vacuum state, then the output state $\Psi _{out}$ = $T$
$\Psi _{in}=\sqrt{\frac 13}[\left| 100\right\rangle +\left| 010\right\rangle
+\left| 001\right\rangle ]_{1^{\prime }2^{\prime }3^{\prime }},$ where, 1',
2', 3' refer to three output ports of tritter. This is the W state which we
want to produce. The probability of success is 100\%. Obviously, besides one
single-photon source and one common tritter, no other element is needed, so
this scheme is very simple.

We can generalize this method to produce an arbitrary path W state of one
photon. What we need is just a N$\times $N lossless multiport fiber beam
splitter. This multiport fiber coupler can be described by a unitary N$%
\times $N matrix, where the matrix elements are the probability amplitudes
for transmission from a certain input port to one of the N output ports. The
relation between N input ports [$a_1,$ $a_2,....a_N]$ and N output ports [$%
b_1,$ $b_2,....b_N]$ can be described by the equation [16]

\begin{equation}
\left( 
\begin{array}{c}
b_1 \\ 
b_2 \\ 
. \\ 
. \\ 
. \\ 
b_N
\end{array}
\right) =\left( 
\begin{array}{c}
M_{11}\text{ }M_{12}\text{ }...\text{ }...\text{ }M_{1N} \\ 
M_{21}\text{ }M_{22}\text{ }...\text{ }...\text{ }M_{2N} \\ 
. \\ 
. \\ 
. \\ 
M_{N1}\text{ }M_{N2}\text{ }...\text{ }...\text{ }M_{NN}
\end{array}
\right) \left( 
\begin{array}{c}
a_1 \\ 
a_2 \\ 
. \\ 
. \\ 
. \\ 
a_N
\end{array}
\right) .
\end{equation}
If the input state is $\Psi _{in}=[1,0,.....0]_{1,2,....N},$ i. e., besides
input 1 is in one photon state, the other N-1 inputs are in vacuum states ,
then the output state $\Psi _{out}=[M_{11},M_{21},....M_{N1}],$ i. e., $\Psi
_{out}=M_{11}\left| 10...0\right\rangle +M_{21}\left| 01...0\right\rangle
+....+M_{N1}\left| 00...1\right\rangle ,$ where, $\left| M_{11}\right|
^2+\left| M_{21}\right| ^2+....+\left| M_{N1}\right| ^2=1.$ So, by choosing
a simple multiport fiber coupler, the kind of arbitrary path W state with
one photon can be produced very easily. The successful probability is 100\%.

Next, we consider how to generate the multiphoton polarization W state. For
application, this kind of W state is more useful. We also take how to
produce the three-photon W state $\frac 1{\sqrt{3}}$[$\left|
HHV\right\rangle +\left| HVH\right\rangle +\left| VHH\right\rangle ]$ as an
example, the generalization to the arbitrary multi-photon W state $\alpha
\left| HH.....HV\right\rangle _N+\beta \left| HH.....VH\right\rangle
_N+.....+\nu \left| VH.....HH\right\rangle _N$ is straightforward, where, $H$
means horizontal linear polarization, $V$ means vertical polarization, $%
\left| \alpha \right| ^2$+$\left| \beta \right| ^2+.....+\left| \nu \right|
^2=1.$ In order to generate the three-photon W state, we need three single
photon sources, which produce $H_1,H_2$ and $V_3$ photons respectively and a
3$\times $3 fiber tritter. Here, we use the symmetry fiber tritter. The
transformation matrix is shown in Eq. (1). We let three photons 1, 2 and 3
enter the inputs 1, 2 and 3 of fiber tritter respectively at the same time.
In outputs of tritter, if we only consider the case in which there is only
one photon in each output, then we can get

\begin{equation}
\lbrack e^{i\frac{2\pi }3}+e^{i\frac{4\pi }3}][\left| HHV\right\rangle
+\left| HVH\right\rangle +\left| VHH\right\rangle ]_{1^{\prime }2^{\prime
}3^{\prime }},
\end{equation}
where, 1', 2' and 3' refer to three outputs of tritter. The probability of
getting the W state is $\frac 19,$ which is larger than the scheme in Ref.
[15]. For producing four-photon W state $\frac 12$[$\left| HHHV\right\rangle
+\left| HHVH\right\rangle +\left| HVHH\right\rangle +\left|
VHHH\right\rangle ]$, we can use a 4$\times $4 canonical quarter to do it.
[16]. By postselection, we can get it with probability 1/16, which is just
few less than 2/27 of Ref.[15], but it is more simpler and more easy to do
it. In principle, we can generate an arbitrary multiphoton W state by same
way. What we need is just a N$\times $N fiber coupler which can be described
by the unitary transformation matrix Eq. (2) and N single photon sources. In
practice, the N$\times $N fiber coupler is standard commercial product, so
the problem is single-photon source. Now, there are some methods to produce
the single photon.[17-19]

In summary, we give two very simple schemes to produce two kinds of W
states. These scheme just need a common commercial multiport fiber coupler
and single photon sources, they are very easy to realize in practice. We can
use this scheme to get the path W state with 100\% probability without
further requirements like postselection. Combining with postselection, we
can get multiphoton polarization W state probabilisticly. Of course, we can
produce the more modes W state from less modes W state or more photons W
state from less photons W state by using the same procedure as entanglement
swapping [20], the disadvantage is that the Bell state measurement is
needed, which makes this scheme difficult in practice.

We thank Prof. H. Imai for his support.

\end{document}